\documentclass[fleqn,10pt]{wlscirep}
\usepackage[utf8]{inputenc}
\usepackage[T1]{fontenc}
\title{Mapping borophene onto graphene: Quasi-exact solutions for guiding potentials in tilted Dirac cones}
\author[1]{R. A. Ng}
\author[2]{A. Wild}
\author[2,3,+]{M. E. Portnoi}
\author[1,*]{R. R. Hartmann}
\affil[1]{Physics Department, De La Salle University, 2401 Taft Avenue, 0922 Manila, Philippines}
\affil[2]{Physics and Astronomy, University of Exeter, Stocker Road, Exeter EX4 4QL, United Kingdom}
\affil[3]{ITMO University, St. Petersburg 197101, Russia}

\affil[*]{Richard.hartmann@dlsu.edu.ph}
\affil[+]{M.E.Portnoi@exeter.ac.uk}

\begin{abstract}
We show that if the solutions to the (2+1)-dimensional massless Dirac equation for a given 1D potential are known, then they can be used to obtain the eigenvalues and eigenfunctions for the same potential, orientated at an arbitrary angle, in a tilted anisotropic 2D Dirac material. This simple set of transformations enables all the exact and quasi-exact solutions associated with 1D quantum wells in graphene to be applied to the confinement problem in tilted Dirac materials such as borophene. We also show that smooth electron waveguides in tilted Dirac materials can be used to manipulate the degree of valley polarization of quasiparticles travelling along a particular direction of the channel. We examine the particular case of the hyperbolic secant potential to model realistic top-gated structures for valleytronic applications.
\end{abstract}

\begin{document}

\flushbottom
\maketitle
%
%
\thispagestyle{empty}

\section*{Introduction}
It can be shown using supersymmetric methods that whenever the Schr\"{o}dinger equation can be solved exactly for a one-dimensional (1D) potential, there exists a corresponding potential for which the two-dimensional (2D) Dirac equation admits exact eigenvalues and eigenfunctions~\cite{cooper1988supersymmetry}. A broad class of quasi-1D potentials can also be solved quasi-exactly by transforming the 2D Dirac equation into the Heun equation~\cite{hartmann2014quasi} or one of its confluent forms~\cite{downing2014one,hartmann2017two,hartmann2020guided}, or via the application of Darboux transformations~\cite{schulze2019darboux2,schulze2019higher,schulze2020arbitrary,schulze2020higher,schulze2021dirac,schulze2021first}. These exact and quasi-exact bound-state solutions have direct applications to electronic waveguides in 2D Dirac materials
~\cite{hartmann2010smooth,hartmann2014quasi,downing2014one,hartmann2017two,hartmann2020bipolar,hartmann2020guided}, such as graphene, where the low-energy spectrum of the charge carriers can be described by a Dirac Hamiltonian~\cite{castro2009electronic}, and the guiding potential can be generated via a top gate~\cite{huard2007transport, ozyilmaz2007electronic,liu2008fabrication, gorbachev2008conductance,williams2011gate,rickhaus2015guiding}. Recent advances in device fabrication, utilizing carbon nanotubes as top gates, has enabled the detection of individual guided modes~\cite{cheng2019guiding}, opening the door to several new classes of devices such as THz emitters~\cite{hartmann2020guided,hartmann2020bipolar}, transistors~\cite{hartmann2010smooth}, and ultrafast electronic switching devices\cite{huang2018new}. These advances in electron waveguide fabrication technology make the need for analytic solutions all the more important, since they are highly useful in: determining device geometry, finding the threshold voltage required to observe a zero-energy mode~\cite{hartmann2010smooth}, calculating the size of the THz pseudo-gap in bipolar waveguides~\cite{hartmann2020bipolar}, as well as ascertaining the optical selection rules~\cite{hartmann2020bipolar,hartmann2020guided} in graphene heterostructures.


In extension to the well-known case of graphene, Dirac cones can in general possess valley-dependent tilt\cite{zabolotskiy2016strain}. There are only a handful of 2D electronic systems that have been predicted to host these tilted cones\cite{katayama2006pressure,goerbig2008tilted,morinari2010topological,lu2016tilted,muechler2016topological,chiu2017type,varykhalov2017tilted,tao2018two,geilhufe2018chemical,polozkov2019carbon}, one of which is $8$-$Pmmn$ borophene\cite{zhou2014semimetallic,zabolotskiy2016strain}, which has attracted considerable attention.
In general, boron-based nanomaterials are a growing field of interest\cite{mannix2018borophene,wang2019review,ou2021emergence}; indeed, exploring the tilt of 2D Dirac cones in the context of $8$-$Pmmn$ borophene has recently led to a plethora of theoretical works spanning many fields of research\cite{verma2017effect,cheng2017anisotropic,islam2017signature,islam2018magnetotransport,jalali2018tilt,zhang2018oblique,singh2018nonlinear,sengupta2018anomalous,yang2018effects,nguyen2018klein,jalali2019polarization,champo2019metal,zare2019negative,ibarra2019dynamical,zhou2019valley,zhang2019anomalous,faraei2019perpendicular,zare2019thermoelectric,herrera2019kubo,sinha2019spin,zhang2019velocity,jafari2019electric,farajollahpour2019solid,paul2019fingerprints,sandoval2020floquet,sengupta2020anomalous,yar2020effects,jalali2020undamped,das2020tunable,zheng2020anisotropic,farajollahpour2020synthetic,rostami2020probing,kapri2020valley,faraei2020electrically,zhou2020anomalous,zhou2020valley,li2020photoinduced,napitu2020photoinduced,li2021novel,mojarro2021optical,kong2021oblique,tan2021anisotropic,lovsic2021effects,xu2021light,fu2021coulomb,jalali2021tilt,pattrawutthiwong2021possible} from optics to transport, and many more. The spectacular rise of borophene, and the growing interest in tilted Dirac materials, has led to the revisiting of several well-known problems in graphene, e.g., Klein tunneling~\cite{nguyen2018klein} and transport across quasi-1D heterostructures~\cite{zhang2018oblique,zhou2019valley,yesilyurt2019electrically,zhou2020valley}, within the context of tilted Dirac materials. As mentioned previously, methods such as supersymmetry and reducing the Dirac equation to the Heun equation, utilize solutions of known problems, to generate solutions to new ones. This begs the question: does a simple mapping exist which would
allow us to harness the large body of exact and quasi-exact
solutions for 1D waveguides in graphene and then apply them to materials with tilted Dirac cones?

In what follows, we show that the differential equations governing guided modes in an anisotropic tilted Dirac waveguide (orientated at an arbitrary angle) can, via a simple transformation, be mapped onto the graphene problem, i.e., transformed into the massless 2D Dirac equation~\cite{castro2009electronic}, for the \textit{same} potential, but of modified strength, effective momentum, and modified energy scale. 
After outlining the transformation, we study the particular case of the hyperbolic secant potential, which in graphene is known to admit quasi-exact solutions to the eigenvalue problem; but nevertheless, the whole spectrum can be obtained via a semi-analytic approach \cite{hartmann2010smooth,hartmann2014quasi}. We use this known graphene waveguide spectrum to generate the corresponding tilted waveguide spectrum, which is verified using a transfer matrix method. Finally, we discuss valleytronic applications.


\section*{Transformation}
\begin{figure}
    \centering
    \includegraphics[width=0.5\textwidth]{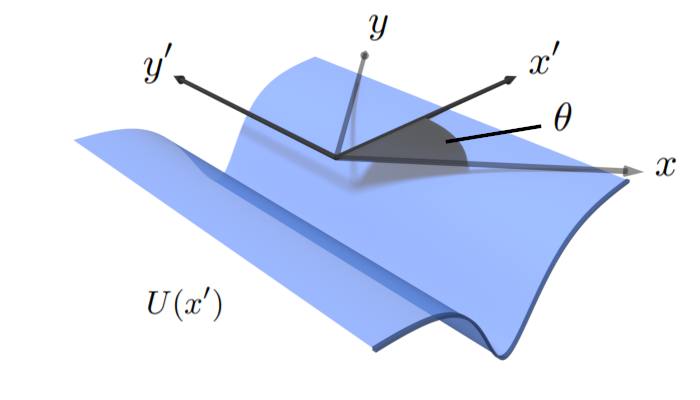}
    \caption{A schematic diagram of an electrostatic potential, $U(x')$, created by an applied top-gate voltage in a tilted 2D Dirac material. The waveguide is orientated at an angle of $\theta$, relative to the $x$-axis of the crystal. The potential is invariant along the $y'$-axis, and varies in strength along the $x'$-axis. The $x'-y'$ axes are denoted by the solid black arrows, whereas the crystallographic axes $x-y$ are shown by the light gray arrows. 
    }
    \label{fig:geometry}
\end{figure}
\begin{figure}
    \centering
    \includegraphics[width=0.8\textwidth]{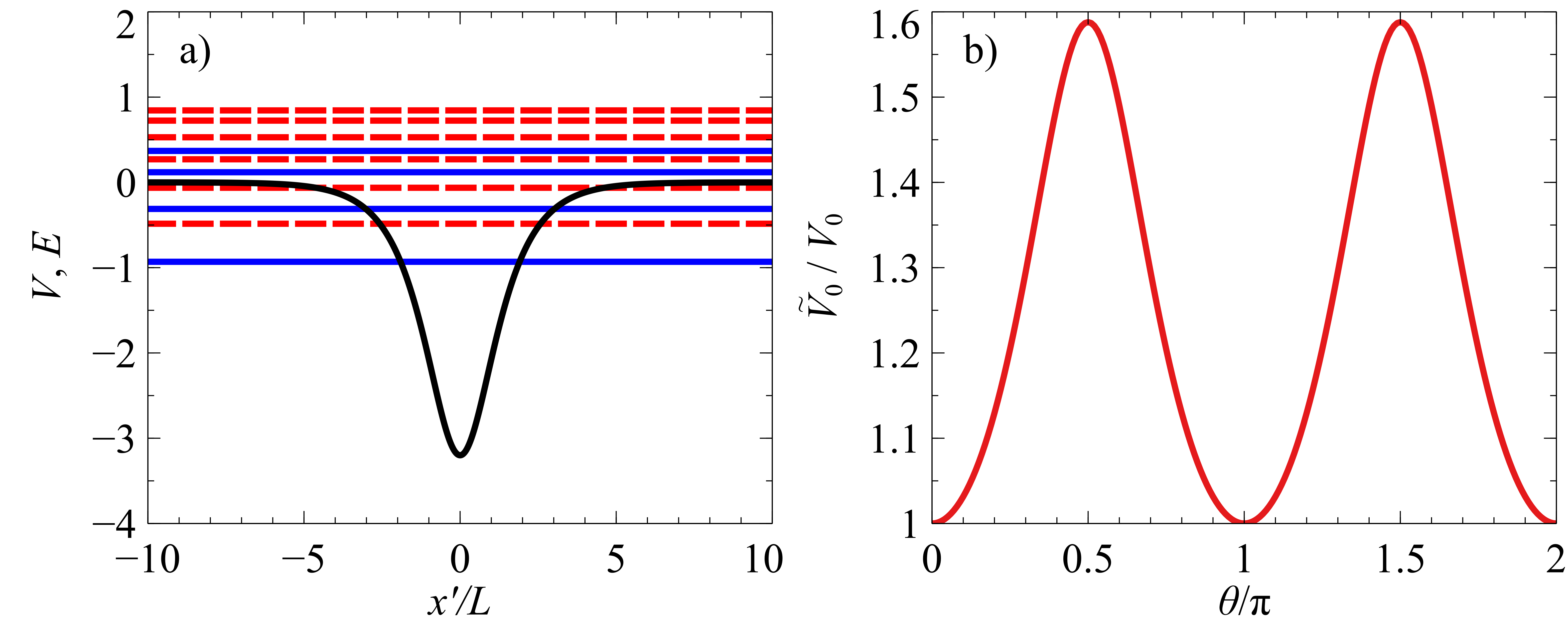}
    \caption{(a) The black curve shows the hyperbolic secant potential $V(x')=-V_0/\cosh(x'/L)$, for the case of $V_0=3.2$. The dashed and solid horizontal lines are the bound-state energy levels for the $s=1$ and $s=-1$ chirality, respectively (corresponding to the $K$ and $K'$ valley in the effective graphene sheet), for the case of $\Delta=k_{y'} L=1$, when the waveguide is orientated at angle $\theta=0$. The tilted Dirac material is defined by parameters $v_x=0.86\,v_{\mathrm{F}}$, $v_y=0.69\,v_{\mathrm{F}}$ and $v_t=0.32\,v_{\mathrm{F}}$. (b) The relative strength of the effective potential's depth, $\widetilde{V}_{0}$, compared to the actual potential depth $V_0$. 
    }
    \label{fig:potential}
\end{figure}

The Hamiltonian describing the guided modes contained within a smooth electron waveguide in a tilted Dirac material can be written as
\begin{equation}
\hat{H}=\hbar\left(v_{x}\sigma_{x}\hat{k}_{x}+sv_{y}\sigma_{y}\hat{k}_{y}+sv_{t}\sigma_{0}\hat{k}_{y}\right)+\sigma_{0}U\left(x,y\right),
\label{eq:Ham_orginal}
\end{equation}
where $\hat{k}_x=-i \partial_x$, $\hat{k}_y=-i \partial_y$, $\sigma_{x,y}$ are the Pauli spin matrices, $\sigma_{0}$ is the identity matrix, $v_{x}$ and $v_{y}$ are the anisotropic velocities, $v_{t}$ is the tilt velocity and $s=\pm1$ is the valley index number; here $s=1$ and $s=-1$ are analogous to the $K$ and $K'$ valley respectively. In what follows, we set $s=1$, but it should be noted that the other valley's eigenvalues can be obtained by replacing $v_y$ and $v_t$ with $-v_y$ and $-v_t$. In general, the crystallographic orientation is not known, nor is it currently possible to deposit the top gate at a selected angle relative to the crystallographic axis. Therefore, we shall solve for the case of a waveguide at an arbitrary angle relative to the crystallographic axis ($x-y$). The electrostatic potential, $U\left(x,y\right)$, is 1D, directed along the $y'$-axis, and varies along the $x'-$axis (see Fig.~\ref{fig:geometry} for geometry), i.e., $U=U\left(x'\right)$. We rotate the $x-y$ axes counterclockwise through an angle $\theta$. The new axes $x'-y'$ are defined by the original coordinates via the transformation:
\begin{align}
    x'&=x\cos\theta+y\sin\theta, 
\nonumber \\ 
    y'&=- x\sin\theta+y\cos\theta.
\end{align}
Hence, the wave vector operators in the non-rotated coordinate system $\hat{k}=\left(\hat{k}_{x},\hat{k}_{y}\right)$ are expressed in the rotated coordinate frame, $\hat{k}'=\left(\hat{k}_{x'},\hat{k}_{y'}\right)$, via the relations:
\begin{align}
\hat{k}_{x}&=\cos\theta \hat{k}_{x'}-\sin\theta \hat{k}_{y'}, \nonumber \\ 
\hat{k}_{y}&=\sin\theta \hat{k}_{x'}+\cos\theta \hat{k}_{y'}.
\end{align}
The Hamiltonian, Eq.~(\ref{eq:Ham_orginal}), acts on the two-component Dirac wavefunction $\Psi=\left(\psi_{A}\left(x'\right),\,\psi_{B}\left(x'\right)\right)^{\intercal} e^{ik_{y'} y'}$ to yield the coupled first-order differential equations $\hat{H} \Psi = \varepsilon \Psi$, where $\psi_{A}$ and $\psi_{B}$ are the wavefunctions associated with the $A$ and $B$ sublattices of the tilted Dirac material. These coupled first-order differential equations can be recast into the same equations used to describe guided modes propagating along a smooth electron waveguide in graphene:
\begin{equation}
\left(
\sigma_{x}\hat{k}_{x'}+\sigma_{y}\widetilde{\Delta}+\sigma_{0}\widetilde{V}\left(x'\right)
\right)
\Phi\left(x'\right)=\widetilde{E}\Phi\left(x'\right),
\end{equation}
where the effective potential $\widetilde{V}$, energy $\widetilde{E}$, and momentum $\widetilde{\Delta}$ are obtained from the original tilted case via the relations:   
\begin{align}
\widetilde{V}\left(x'\right)&=\frac{lV\left(x'\right)}{l^{2}-t^{2}\sin^{2}\theta},
\nonumber \\
\widetilde{E}&=\frac{l}{l^{2}-t^{2}\sin^{2}\theta}\left(E-\frac{t\cos\theta}{l^{2}}\Delta\right),
\nonumber \\
\widetilde{\Delta}&=\frac{T\Delta}{l\sqrt{l^{2}-t^{2}\sin^{2}\theta}},
\label{eq:mapping}
\end{align}
where $V\left(x'\right)= U\left(x'\right) L/\hbar v_x$, $E=\varepsilon L/\hbar v_x$ and $\Delta=k_y'L$, and $L$ is a constant, associated with the effective width of the potential. We define the tilt and anisotropy parameters as $t=v_{t}/v_{x}$ and $T=v_{y}/v_{x}$, respectively, and $l=\sqrt{1-\left(1-T^{2}\right)\sin^{2}\theta}$. The eigenfunctions of the guided modes in the effective graphene sheet, $\Phi$, can be mapped onto the tilted Dirac spinor components, $\psi_{A}$ and $\psi_{B}$, via the expression:
\begin{equation}
\Phi\left(x'\right)=\left(\begin{array}{c}
\left(1+\mu\right)\psi_{A}+\left(1-\mu\right)\psi_{B}e^{-i\varphi}\\
\left(1-\mu\right)\psi_{A}+\left(1+\mu\right)\psi_{B}e^{-i\varphi}
\end{array}\right)
e^{-i\sin\theta
{\displaystyle \int}
\frac{t
\left(V-E\right)
+k_{y'}
\left(1+t^{2}-T^{2}\right)
\cos\theta}{l^{2}-t^{2}\sin^{2}\theta}dx'}
\end{equation}
where $\varphi=\arctan\left(T\tan\theta\right)$ and $\mu=\left(l-t\sin\theta\right)^{\frac{1}{2}}\left(l+t\sin\theta\right)^{-\frac{1}{2}}$. It then follows that if the eigenfunctions and eigenvalues are known for the potential $\widetilde{V}$ in graphene, one can immediately write down the eigenfunctions and eigenvalues of a 1D confining potential of the same form, orientated at an arbitrary angle, in a tilted Dirac material. Conversely, if a quasi-1D potential readily admits exact or quasi-exact solutions for the tilted case, and no solutions are known for the graphene problem, then our mapping method can be used to obtain the eigenfunctions and eigenvalues for the case of graphene. This mapping also reveals the angular dependence of the number of guided modes contained within the waveguide. Namely, it can be seen from Eq.~(\ref{eq:mapping}) that rotating the orientation of the waveguide is equivalent to varying the effective depth of the potential (see Fig.~\ref{fig:potential}b). Indeed, the effective potential's depth, $\widetilde{V}_{0}$, is equal to the actual potential's depth, $V_0$ at $\theta=0$ and rises to a maximum value of $\widetilde{V}_0 / V_0= T/(T^{2}-t^{2})$ at $\theta=\pi/2$.

It should be noted that to perform the same transformations for the other chirality (i.e., the $s=-1$, or graphene $K'$ valley analog), one must exchange $t$ with $-t$, and $T$ with $-T$. Therefore, the eigenvalue spectrum of one valley can be obtained by a reflection of the other valley's eigenvalue spectrum about the $k_{y'}$-axis. It can be seen from Eq.~(\ref{eq:mapping}) that in the absence of a tilt term, i.e., $t=0$, the eigenvalue spectrum of a given valley is symmetric with respect to $k_{y'}$. Thus, both chiralities have the same band structure. Similarly for $t \neq 0$, if the waveguide is orientated such that $\cos{\theta}=0$, the eigenvalue spectrum is chirality-independent. In all other cases, the energy spectrum for a given valley lacks $E(k_{y'})=E(-k_{y'})$ symmetry. This gives rise to the possibility of utilizing smooth electron waveguides in tilted Dirac materials as the basis of valleytronic devices. This will be discussed in the penultimate section.

\section*{Quasi-exact solution to the tilted Dirac equation for the hyperbolic secant potential} 
\begin{figure}
    \centering
    \includegraphics[width=0.75\textwidth]{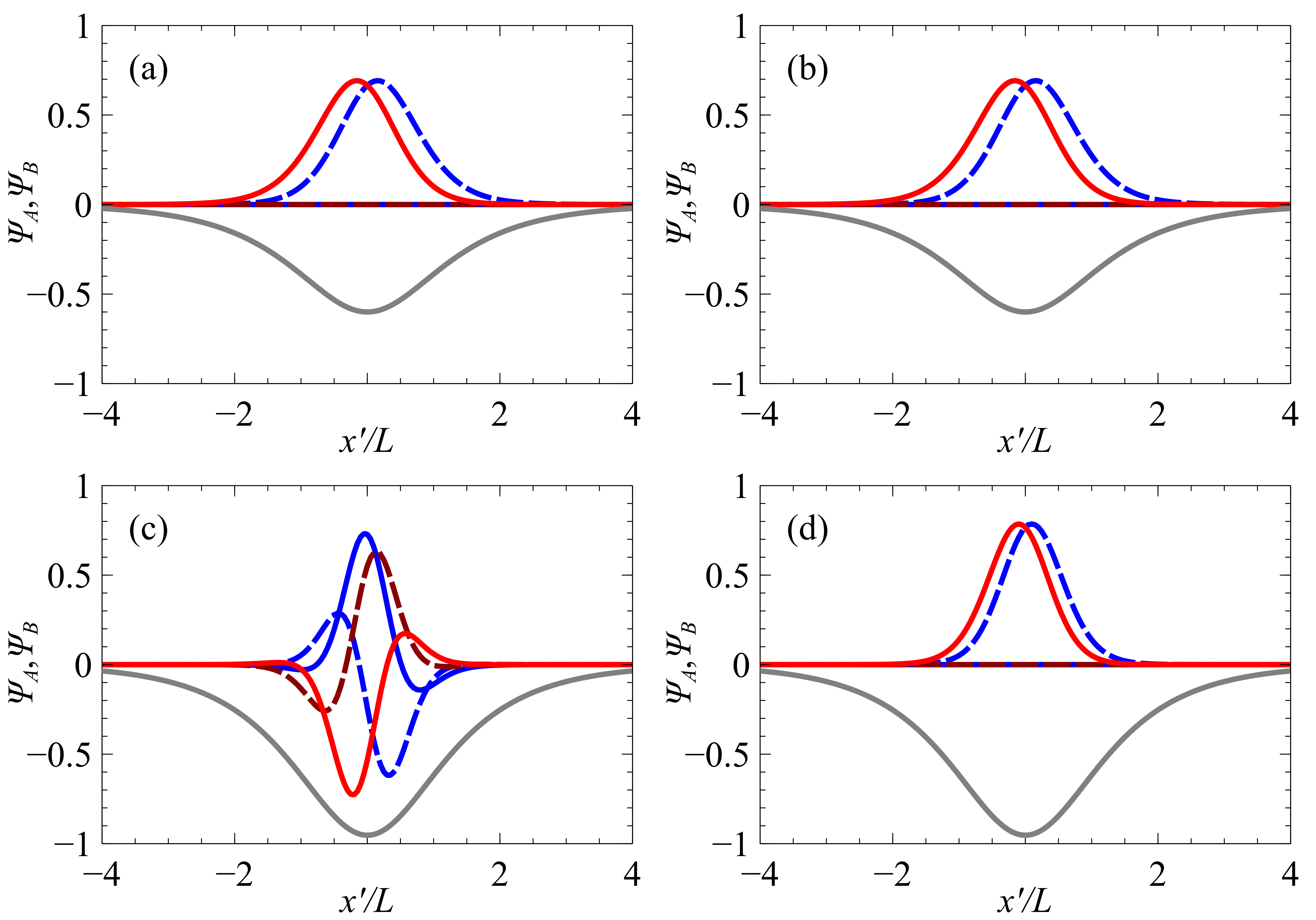}
    \caption{The normalized zero-energy state (lowest positive momentum) wavefunctions of the hyperbolic secant potential, $V(x')=-V_0/\cosh(x'/L)$, of strength $V_0=-3.2$ and orientation: (a) $\theta=0$ and (c) $\theta=\pi/2$ in a tilted Dirac material defined by parameters $v_x=0.86\,v_{\mathrm{F}}$, $v_y=0.69\,v_{\mathrm{F}}$ and $v_t=0.32\,v_{\mathrm{F}}$, for the $s=1$ chirality, i.e., the $K$ valley in the effective graphene sheet. The corresponding wavefunctions of the effective graphene waveguide are given in panels (b) and (d). The solid red and blue lines correspond to real part of $\psi_A$ and $\psi_B$ respectively, while the dashed lines correspond to their imaginary parts. The grey line shows the potential as a guide to the eye.}
    \label{fig:wavefunctions}
\end{figure}

\begin{figure}
    \centering
    \includegraphics[width=0.7\textwidth]{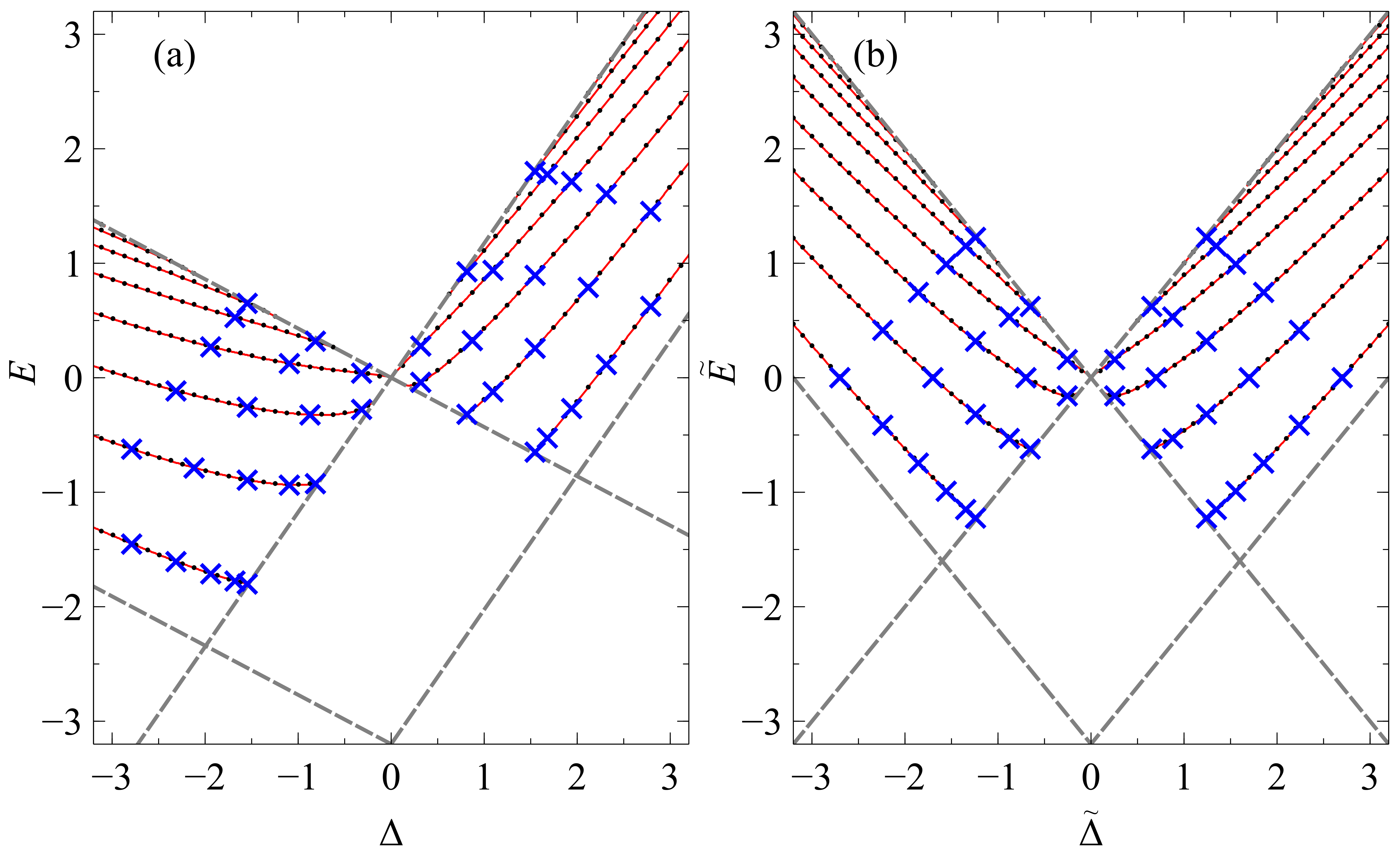}
    \includegraphics[width=0.7\textwidth]{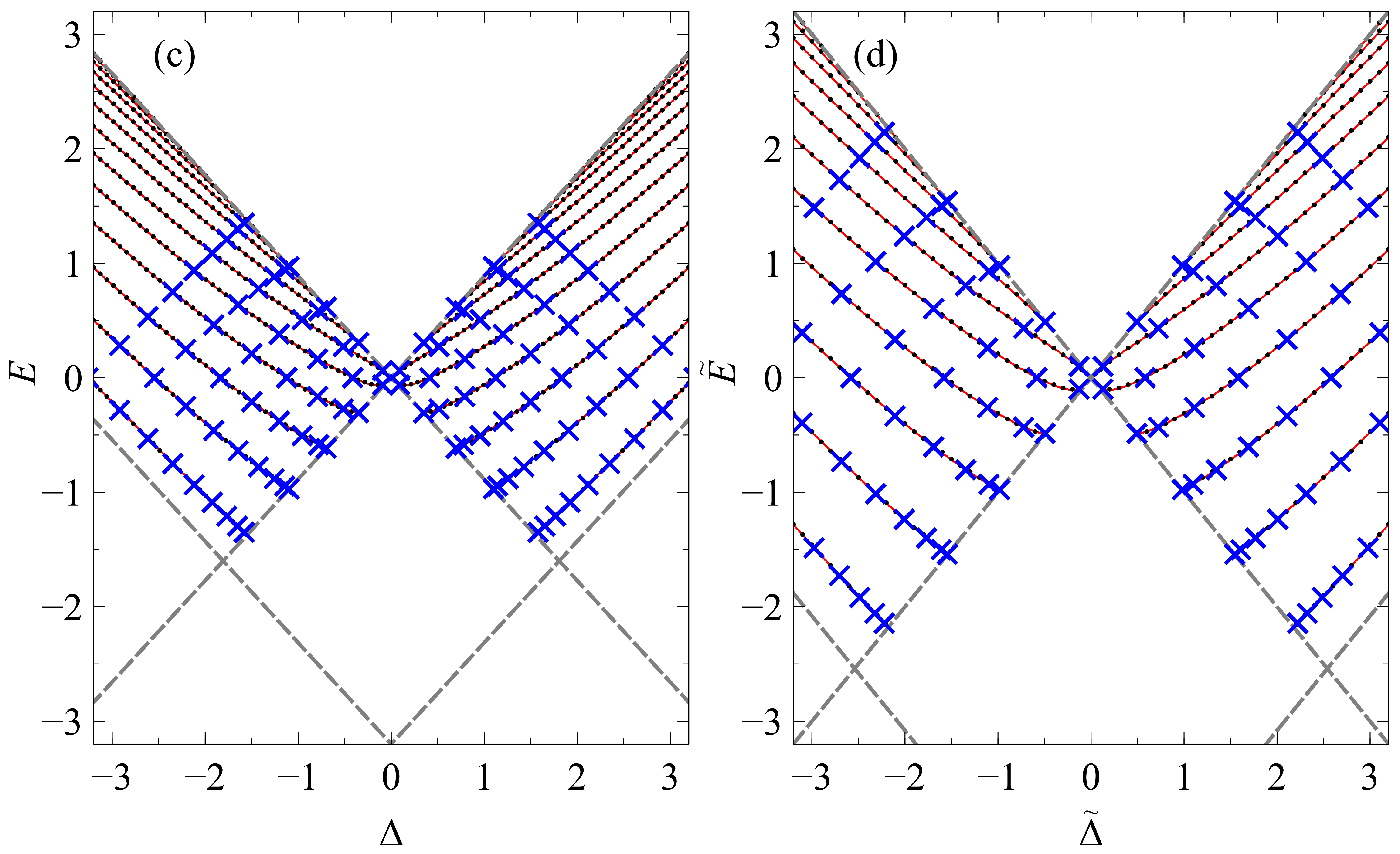}
    \caption{The energy spectrum of confined states (for the $s=1$ chirality, i.e., the $K$ valley in the effective graphene sheet) in the hyperbolic secant potential $V(x')=-V_0/\cosh(x'/L)$, of strength $V_0=3.2$, as a function of dimensionless momentum along the waveguide, $\Delta=k_y'L$, for the orientations (a) $\theta=0$ and (c) $\theta=\pi/2$ in a tilted Dirac material, defined by parameters $v_x=0.86\,v_{\mathrm{F}}$, $v_y=0.69\,v_{\mathrm{F}}$ and $v_t=0.32\,v_{\mathrm{F}}$. The energy spectra of the effective graphene waveguide from whence they came, are given in panels (b) and (d), respectively. The black dots denote the semi-analytic eigenvalues, the blue crosses represent the quasi-exact eigenvalues, and the solid red lines show the eigenvalues numerically obtained via a transfer matrix method. The boundary at which the bound states merge with the continuum is denoted by the grey dashed lines.}
    \label{fig:dispersion}
\end{figure}

In this section, we shall apply our simple transformations, given in Eq.~(\ref{eq:mapping}), to generate the energy spectrum of a smooth electron waveguide in a tilted Dirac material for a potential which has been studied in depth in graphene:
\begin{equation}
    V\left(x'\right)=-\frac{V_0}{\cosh\left(x'/L\right)}.
    \label{eq:potential}
\end{equation}
This potential, shown in Fig.~\ref{fig:potential}a, belongs to the class of quantum models which are quasi-exactly solvable~\cite{turbiner1988guantum,bender1998quasi,downing2013solution,hartmann2014quasi,hartmann2014bound,hartmann2016exciton,hartmann2017pair,ushveridze2017quasi}, where only some of the eigenfunctions and eigenvalues are found explicitly. The depth of the well is given by $V_0$, and the potential width is characterized by the parameter $L$. Here $V_0$ and $L$ are taken to be positive parameters. In graphene, the wavefunctions can be solved in terms of Heun polynomials, which reduce to hypergeometric functions for the case of zero energy~\cite{hartmann2010smooth,hartmann2014quasi}. For zero-energy modes ($\widetilde{E}=0$), the permissible values of $\widetilde{\Delta}$ are given by the simple relation $\widetilde{\Delta}=\widetilde{V}_0-n-\frac{1}{2}$, where $\widetilde{V}_0$ is the depth of the effective potential and $n$ is a non-negative integer.
For non-zero energies, exact energy eigenvalues can be obtained when the Heun polynomials are terminated~\cite{hartmann2014quasi}. To illustrate the power of our mapping method, we apply the transformations given in Eq.~(\ref{eq:mapping}) to the well-known zero-energy solutions to the 2D Dirac equation for the 1D hyperbolic secant potential. The corresponding tilted Dirac equation solutions become:
\begin{equation}
E=\frac{t\cos\theta}{T\sqrt{l^{2}-t^{2}\sin^{2}\theta}}\left[V_{0}-\left(n+\frac{1}{2}\right)\frac{l^{2}-t^{2}\sin^{2}\theta}{l}\right],
\label{eq:E_0_mapped}
\end{equation}
and their corresponding $n=0$ eigenfunctions for two different waveguide orientations are shown in Fig.~\ref{fig:wavefunctions}, for the case of borophene, i.e., $v_x=0.86\,v_{\mathrm{F}}$, $v_y=0.69\,v_{\mathrm{F}}$ and $v_t=0.32\,v_{\mathrm{F}}$. In Fig.~\ref{fig:dispersion} we plot the borophene eigenvalue spectrum for the potential defined by $V_0=3.2$ for two orientations, $\theta=0$ and $\theta=\pi/2$, as well as the graphene waveguide spectra (quasi-analytically determined~\cite{hartmann2014quasi}) used in the mapping. In the same figure, we also plot the numerical solutions to the tilted Dirac problem obtained via a transfer matrix method (see supplementary material). We show in blue crosses the exact solutions given in Eq.~(\ref{eq:E_0_mapped}) together with the complete set of mapped quasi-exact solutions given in ref.~\cite{hartmann2014quasi}. It can be seen from Fig.~\ref{fig:dispersion} that the waveguide orientated at $\theta=\pi/2$ contains more bound states than the waveguide orientated at $\theta=0$. This is a result of the effective graphene potential being deeper, and thus supporting more guided modes. It should be noted that for potentials which vanish at infinity, i.e., $V(\pm\infty)=E=0$, only the zero-energy modes are truly confined, since the density of states vanishes outside of the well. Guided modes occurring at non-zero energies can always couple to continuum states outside of the well, thus having a finite lifetime.

\section*{Valleytronic applications} 
\begin{figure}[ht]
    \centering
    \includegraphics[width=0.4\textwidth]{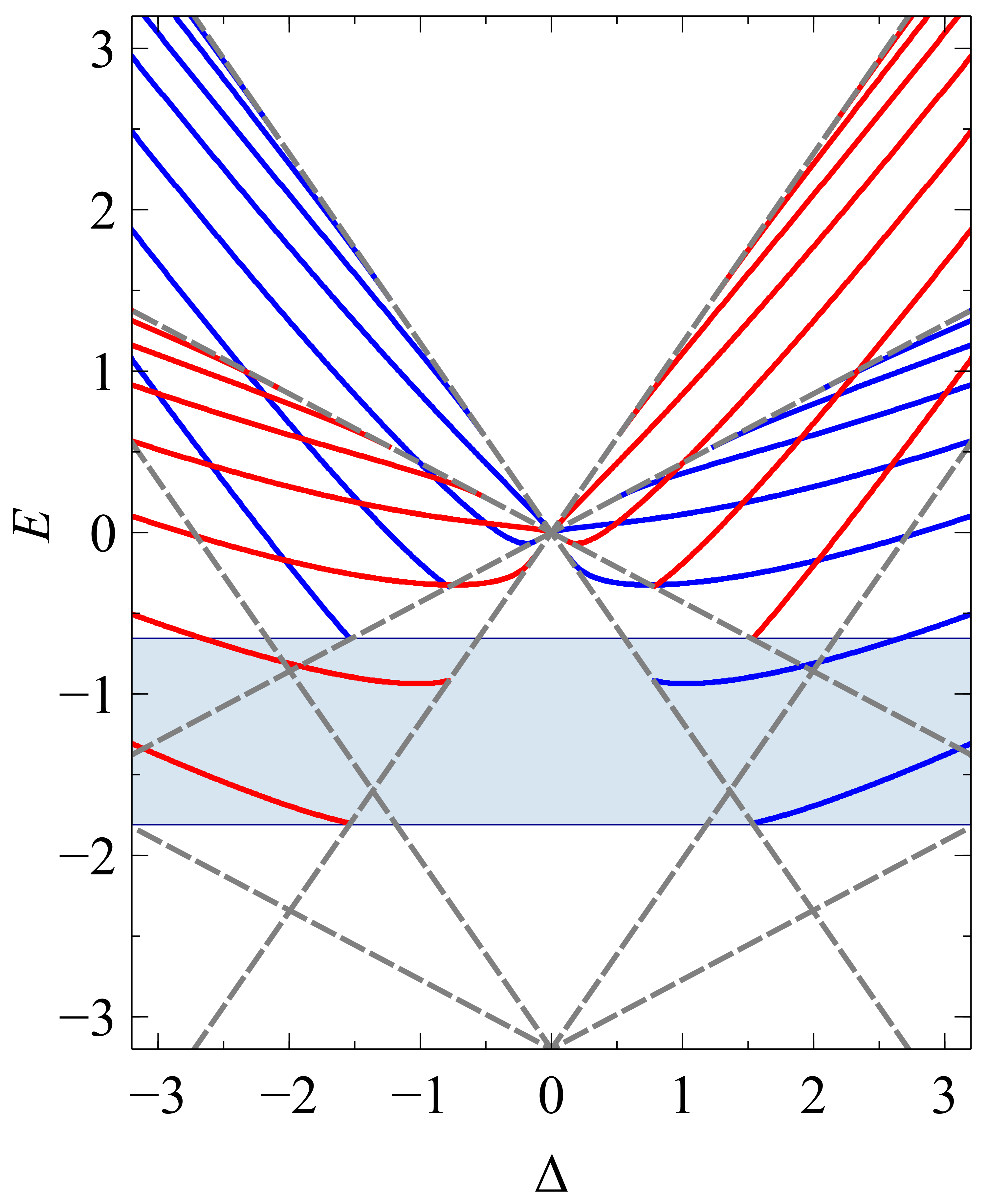}
    \caption{The energy spectrum of confined states in the hyperbolic secant potential $V(x')=-V_0/\cosh(x')$, of strength $V_0=3.2$, as a function of $\Delta$ for the $s=1$ and $s=-1$ chirality (i.e., the $K$ and $K'$ valley in the effective graphene sheet), depicted by the red and blue lines, respectively. The waveguide is orientated at $\theta=0$ and the tilted Dirac material is defined by parameters $v_x=0.86\,v_{\mathrm{F}}$, $v_y=0.69\,v_{\mathrm{F}}$ and $v_t=0.32\,v_{\mathrm{F}}$. The boundary at which the bound states merge with the continuum is denoted by the grey dashed lines. The shaded area represents the energy range for which full valley polarization can be achieved for a given value of $\Delta$.}
    \label{fig:valley}
\end{figure}

It has been suggested that the valley quantum number can be used as a basis for carrying information in graphene-based devices~\cite{rycerz2007valley} in an analogous manner to spin in semiconductor spintronics \cite{vzutic2004spintronics,vitale2018valleytronics}. Unlike in the case of graphene (in the conical regime), smooth electrostatic potentials in tilted Dirac materials can be utilized as a means to achieve valley polarization. 
The majority of studies have focused on tunneling across electrostatically-induced potential barriers, and valley filtering and beam splitting have been demonstrated~\cite{nguyen2018klein,zhang2018oblique,zhou2019valley,yesilyurt2019electrically,zhou2020valley}. It has also been shown that the allowed transmission angles through a potential can be controlled using magnetic barriers~\cite{yesilyurt2019electrically}. We propose a change in geometry: rather than studying chirality-dependent transmission across barriers, we shift the focus to studying guided modes along quasi-1D confining potentials. The conductance along such a channel can be measured by placing one terminal at each end. According to the Landauer formula, when the Fermi level is set to energy $E$ (by modulating the back-gate voltage~\cite{cheng2019guiding}), the conductance along the waveguide is simply $2(n_K+n_{K'})e^{2}/h$, where $n_K$ and $n_{K'}$ are the number of modes belonging to the $s=1$ and $s=-1$ chirality, respectively (or the $K$ and $K'$ valley in the effective graphene sheet), at that particular energy.


In a 2D Dirac material subject to a quasi-1D potential, the introduction of the tilt parameter breaks the $E(\Delta)=E(-\Delta)$ symmetry for a given valley. Indeed, for a given valley, the additional tilt term increases the particle velocity along one direction of the barrier and decreases it along the opposite direction; and vice versa for the other valley. For the case of type-I Dirac materials, i.e., $t<T$, it can be seen from Fig.~\ref{fig:valley} that for a given sign of $\Delta$, the eigenvalues of the critical solutions (sometimes referred to as the zero-momentum solutions: i.e., bound states with energy $|E|=|\Delta|$)
belonging to the two valleys are different. Thus, providing that $t\neq0$ and $\cos{\theta}\ne0$, there exists a range of energies for which there will be more bound states propagating along a particular direction belonging to one valley than the other, i.e., valley polarization. The degree of valley polarization can be controlled by varying the strength of the electrostatic potential and by changing the position of the Fermi level, which in practical devices is achieved by modulating the back-gate voltage\cite{cheng2019guiding}. Full valley polarization can be achieved for energies less than the lowest lying supercritical state (defined as a bound state with energy $E = -\Delta$) belonging to the valley where the tilt term enhances the particle's velocity, indicated by the shaded region in Fig.~\ref{fig:valley}. 

For type-III tilted Dirac materials, i.e., $t=T$, full valley polarization occurs for all energies and all orientations of the waveguide. For such materials the infinite number of positive-energy critical solutions of graphene map onto the zero-energy modes of a type-III Dirac material. Consequently, the infinitely many zero-energy modes will give rise to a sharp peak in the conductance along the channel when the Fermi energy is in the proximity of $E=0$. This is in stark contrast to the case of graphene, where the creation of an infinite number of zero-energy states requires an infinitely deep and wide potential~\cite{clemence1989low,lin1999levinson,calogeracos2004strong, ma2006levinson,hartmann2017two}. Since the potential possesses infinitely many bound states, which infinitely accumulate at $E=0$, a type-III Dirac material could be used as a THz emitter; namely, the Fermi level could be set below $E=0$, and optical photons can be absorbed from low-lying energy levels to $E=0$. Then the photo-excited carriers can relax back down to the Fermi level via the emission of THz photons through the closely spaced energy levels in the proximity of zero energy.

Lastly, for type-II tilted Dirac materials, i.e., $t>T$, full valley polarization occurs for all energies; however, bound states occur only for orientations within the range $-1/\sqrt{t^{2}-T^{2}}<\tan\left(\theta+n\pi\right)<1/\sqrt{t^{2}-T^{2}}$. In the limit at which $\theta$ becomes imaginary, i.e., the boundary at which the equi-energy surfaces become unbounded, the effective potential required for mapping diverges.

\section*{Conclusion}
We have shown that if the eigenvalues and eigenfunctions of a quasi-1D potential in graphene are known, then they can be used to obtain the corresponding results for the same potential (with modified strength, orientated at arbitrary angle), for an anisotropic, tilted 2D Dirac material. Therefore, all the rich physics associated with guiding potentials in graphene, e.g., THz pseudo-gaps in bipolar waveguides, can be revisited in the context of tilted Dirac materials, but with the distinct advantage of knowing the eigenvalues and eigenfunctions. We have also shown that in stark contrast to smooth electron waveguides in graphene, valley degeneracy can be broken in tilted 2D Dirac materials for a broad range of waveguide orientations, anisotropy and tilt parameters. The degree of valley polarization along the waveguide can be controlled by varying the potential strength of the top gate, and also by changing the back-gate voltage. Tilted 2D Dirac materials, such as borophene, are therefore promising building blocks for tunable valleytronic devices.

\section*{Methods}
The Supplementary Information contains a full description of the transfer matrix method used to calculate the band structure of guiding potentials in 2D Dirac materials with tilted Dirac cones. 

\section*{Data Availability}
This study did not generate any new data.

\bibliography{sample}

\section*{Acknowledgements}
This work was supported by the EU H2020 RISE projects TERASSE (H2020-823878) and DiSeTCom (H2020-823728). RAN is funded by the DOST-SEI ASTHRDP program. AW is supported by a UK EPSRC PhD studentship (Ref. 2239575) and by the NATO Science for Peace and Security project NATO.SPS.MYP.G5860. RRH acknowledges financial support from URCO (14 F 1TAY20-1TAY21). The work of MEP was supported by the Russian Science Foundation (Project No. 20-12-00224).

\section*{Author contributions statement}
All authors contributed equally to the writing of the manuscript.

\section*{Additional information}
\textbf{Competing interests} The authors declare that they have no competing interests.

\end{document}


\renewcommand{\theequation}{S\arabic{equation}}
\renewcommand{\thefigure}{S\arabic{figure}}
\flushbottom
\maketitle

\thispagestyle{empty}

\section*{Numerical Model of a Waveguide in a 2D Dirac Material with Tilted Dirac Cones}

In order to confirm the results of the main text, we discretize smooth waveguides and numerically calculate their band structure using the transfer matrix method. We begin by determining the wavefunction in a single 1D square well for a Dirac material with tilted cones, as an extension to the graphene case\cite{GrapheneWell}. An arbitrary number of square wells are placed one after another to approximate the smooth waveguides discussed in the main text. For the system of sequential wells we set up a transfer matrix which links the wavefunction components in the leftmost and rightmost well. Bound states yielding the band structure are found numerically by placing boundary conditions on the transfer matrix. The boundary conditions ensure the wavefunction has plane-wave solutions inside the waveguide and decaying solutions at the edges of the system.

\subsection*{Solution to the Dirac Equation with Tilted Dirac Cones in a Square Well}
The Dirac equation for electrons in a 2D semimetal with tilted Dirac cones subjected to the potential $U(x,y)$ has the form 
\begin{equation}
    v_x \big( s t \sigma_0 \hat{p}_y + s T \sigma_y \hat{p}_y + \sigma_x \hat{p}_x \big) \Psi(x,y) = \big( \epsilon - U(x,y) \big) \Psi(x,y),
\end{equation} with energy $\epsilon$, momentum $\hat{p}_i$ where $i=x,y$, Fermi velocity $v_x$, tilt parameter $t$, anisotropy parameter $T$ and valley index $s = \pm 1$. The identity matrix is $\sigma_0$, the Pauli matrices are $\sigma_x$ and $\sigma_y$, and the spinor wavefunction is $\Psi(x,y) = ( \phi^\text{A}(x,y), \phi^\text{B}(x,y) )^\intercal$ which is written in the basis of the Bloch sums of the Dirac material. We wish to find the wavefunction of the $j^\text{th}$ well where $j=1,2,...N$. Each well has finite depth $U_j$ with arbitrary width along the $x'$-axis ($x'_{j-1} \leq x' \leq x'_j$) and is invariant along the perpendicular $y'$-axis (see Fig.\,1 of the main text). Next we perform a coordinate transformation on the momentum operators $\hat{p}_x = \hat{p}_{x'}\cos\theta - \hat{p}_{y'}\sin\theta$ and $\hat{p}_y = \hat{p}_{x'}\sin\theta + \hat{p}_{y'}\cos\theta$ before inserting their definition $\hat{p}_{i'} = -i \hbar \partial_{i'}$, where $i'=x',y'$. Due to the invariance along the $y'$-axis we can insert a plane-wave solution along this axis $\Psi_j(x',y') = \Psi_j(x')e^{i k_{y'}y'}$. It is at this point that we introduce an arbitrary length scale $L$ allowing us to define the dimensionless parameters $\xi = x'/L$, $V_j = U_j L/\hbar v_x$, $E = \epsilon L/\hbar v_x$ and $\Delta = k_{y'}L$. Let the wavefunction in the $j^\text{th}$ well be of the form $\Psi_j(\xi) = ( \phi^{\text{A}}_j, \phi^{\text{B}}_j )^{\intercal} e^{i b_j \xi}$, where $b_j$ is in general a complex number. All preceding steps yield the eigenvalue problem \begin{equation}
\label{eq:EigenvalueProblem}
   \begin{pmatrix}
   st\big( b_j \sin\theta + \Delta \cos\theta \big) - \big( E - V_j\big) & b_j\cos\theta - \Delta \sin\theta - isT\big( b_j \sin\theta + \Delta \cos\theta \big) \\ b_j\cos\theta - \Delta \sin(\theta) + isT\big( b_j \sin\theta + \Delta \cos\theta \big) &  st\big( b_j \sin\theta + \Delta \cos\theta \big) - \big( E - V_j\big)
   \end{pmatrix} 
   \begin{pmatrix}
   \phi_{j,\text{A}}\\\phi_{j,\text{B}}
   \end{pmatrix} = \textbf{0},
\end{equation} where $\textbf{0}$ is the null vector. We define the above eigenvalue problem as $[\mathcal{H}_j - \sigma_0 (E - V_j)]( \phi^{\text{A}}_j, \phi^{\text{B}}_j )^\intercal = \textbf{0}$, where for non-trivial solutions we must ensure $\mid \!\! \mathcal{H}_j - \sigma_0 (E - V_j) \!\! \mid = 0$. This step yields two solutions ($\pm$) for the complex parameter \begin{equation}
    \label{eq:b} b_{j,\pm} = \frac{\big[st(V_j - E) + (1 + t^2 - T^2)\Delta \cos\theta\big]\sin\theta \pm \sqrt{(t^2 - T^2)\Delta^2 + 2st\Delta(V_j - E)\cos\theta + l^2(V_j - E)^2}}{l^2 - t^2\sin^2\theta},
\end{equation} where $l = \sqrt{1 - (1 - T^2)\sin^2\theta}$. The first term of Eq.\,(\ref{eq:b}) is always real whilst the second term can be either real or imaginary depending on the parameters. We shall now introduce the subscript $\pm$ corresponding to the two roots of $b_{j,\pm}$. Inserting these solutions back into Eq.\,(\ref{eq:EigenvalueProblem}) allows us to define $\phi^\text{B}_{j,\pm} = \Lambda_{j,\pm} \phi^\text{A}_{j,\pm}$ where
\begin{equation}
    \Lambda_{j,\pm} = \frac{-st \big( b_{j,\pm} \sin\theta + \Delta \cos\theta \big) + (E - V_j)}{b_{j,\pm}\cos\theta - \Delta \sin\theta - isT \big( b_{j,\pm} \sin\theta + \Delta \cos\theta \big)}.
\end{equation} 
The wavefunction in the region $\xi_{j-1} \leq \xi \leq \xi_j$, where $\xi_j = x'_j/L$ can be written as a linear combination of the forward and backward propagating plane-wave solutions: 
\begin{equation}
    \Psi_j(\xi) = C_j \Bigg[ \phi^\text{A}_{j,+} \begin{pmatrix}
        1\\ \Lambda_{j,+}
    \end{pmatrix}e^{i b_{j,+} \xi} + \phi^\text{A}_{j,-} \begin{pmatrix}
        1\\ \Lambda_{j,-}
    \end{pmatrix}e^{i b_{j,-} \xi} \Bigg],
\end{equation}
where $C_j$ is a normalization factor. This can be written as $\Psi_j(\xi) = \Omega_j(\xi)(\alpha_j, \beta_j)^\intercal$ where \begin{equation}
    \Omega_j(\xi) = 
    \begin{pmatrix}
    e^{ib_{j,+}\xi} & e^{i b_{j,-}\xi}\\
    \Lambda_{j,+} e^{ib_{j,+}\xi} & \Lambda_{j,-} e^{ib_{j,-}\xi}
    \end{pmatrix},
\end{equation} 
where $\alpha_j=C_j \phi^\text{A}_{j,+}$ and  $\beta_j=C_j \phi^\text{A}_{j,-}$.

\subsection*{Transfer Matrix}

\begin{figure}
    \centering
    \includegraphics[width=1\textwidth]{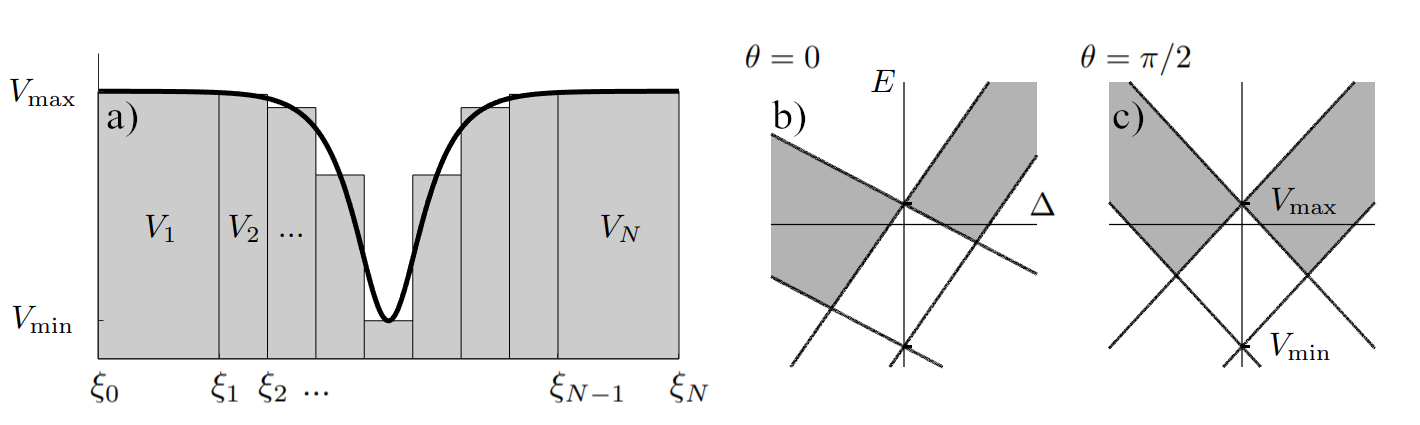}
    \caption{(a) Schematic of the waveguide $V(\xi) \propto -1/\cosh(\xi)$ approximated by $N$ square wells. The depth of the $j^\text{th}$ well is $V_j$ and the leftmost and rightmost wells ($V_1$ and $V_N$) are arbitrarily wide as it is assumed $\xi_0 \to -\infty$ and $\xi_N \to \infty$. The waveguide has a depth of $V_\text{min}$ and will terminate at the edges with a potential of $V_\text{max}$. Panels (b) and (c) represent the regions of wavevectors and energies where the waveguide has bound states in valley $s=1$, for waveguide orientations $\theta = 0$ and $\pi/2$ respectively, when $\mid \!\! T \!\!\mid \neq 1$ and $0 < \mid \!\! t \!\! \mid < \mid \!\! T \!\!\mid$. The results for the second valley $s=-1$ can be found from the $s=1$ case by replacing $\Delta$ with $-\Delta$.}
    \label{fig:WaveguideSchem}
\end{figure}

We now consider a smooth waveguide approximated by $N$ square wells as sketched in Fig.\,\ref{fig:WaveguideSchem}a. The analytics will be limited to a simple waveguide that terminates at the same maximum potential to the left and right, i.e., $V_1 = V_N = V_\text{max}$, but the analysis can be readily extended to the general case. We must first consider our conditions for a bound state. From the perspective of single wells, our bound state must have a plane-wave solution in the deepest well, $V_\text{min}$, which means that in this region, $\text{Im}(b_{j,\pm}) = 0$. Additionally, we must ensure that the wavefunction decays in the leftmost and rightmost wells, $V_\text{max}$, meaning $\text{Im}(b_{1,\pm}) = \text{Im}(b_{N,\pm}) \neq 0$ so that it disappears as $\xi \to \pm \infty$. These two criteria are satisfied by the condition \begin{equation}
\label{eq:Constraints1}
    \{E < V_\text{min} + f_{1} \Delta + f_{2} \! \mid \!\! \Delta \!\! \mid \:\: \cup \:\: E > V_\text{min} + f_{1} \Delta - f_{2} \! \mid \!\! \Delta \!\! \mid \} \:\:\:\: \cap \:\:\:\: \{E > V_\text{max} + f_1 \Delta + f_2 \! \mid \!\! \Delta \!\! \mid \:\: \cap \:\: E < V_\text{max} + f_1 \Delta - f_2 \! \mid \!\! \Delta \!\! \mid \},
\end{equation} where \begin{equation}
\label{eq:Constraints2a}
    f_{1} = \frac{st \!\cos\theta}{l^2},
\end{equation} and \begin{equation}  
    \label{eq:Constraints2b}
    f_{2} = -\frac{\! \mid \!\! T \!\! \mid \!\! \! \sqrt{l^2 - t^2 \sin^2\theta}}{l^2}.
\end{equation} For the reader's convenience these regions have been sketched for two waveguide orientations (see Fig.\,\ref{fig:WaveguideSchem}b and c).

Now that we have established the requirements for bound states in the waveguide, we need to match the wavefunctions of each square well at their boundaries. If we begin on the right side of the waveguide, we must match the wavefunctions $\Psi_j(\xi)$ in regions $j=N-1$ and $j=N$ at the interface $\xi_{N-1}$. This will allow us to write the wavefunction components of region $N$ as a function of the components in region $N-1$: \begin{equation}
        \begin{pmatrix}
        \alpha_N\\
        \beta_N
        \end{pmatrix} = \Omega^{-1}_N(\xi_{N-1})\Omega_{N-1}(\xi_{N-1})\begin{pmatrix}
        \alpha_{N-1}\\
        \beta_{N-1}
        \end{pmatrix}.
\end{equation} Iterating this process will allow us to write the components of the wavefunctions in the right region ($j=N$) as a function of the components of the left region ($j=1$) as $(\alpha_1, \beta_1)^\intercal = \textbf{T} (\alpha_N, \beta_N)^\intercal$, where the transfer matrix is defined as \begin{equation}
    \textbf{T} = \begin{pmatrix}
    T_{\alpha \alpha} & T_{\alpha \beta} \\ T_{\beta \alpha} & T_{\beta \beta}
    \end{pmatrix} = \prod_{j=1}^{N-1} \Omega_{j}^{-1}(\xi_{j}) \Omega_{j+1}(\xi_{j}).
\end{equation} 
To ensure the wavefunction decays at $\xi \to \pm \infty$ we set $\alpha_1 = 0$ and $\beta_N = 0$, hence \begin{equation}
    \begin{pmatrix}
        0\\
        \beta_1
    \end{pmatrix}
    = \begin{pmatrix}
    T_{\alpha \alpha} & T_{\alpha \beta} \\ T_{\beta \alpha} & T_{\beta \beta}
    \end{pmatrix} \begin{pmatrix}
        \alpha_N\\
        0
    \end{pmatrix}.
\end{equation} This places a constraint on the transfer matrix $T_{\alpha \alpha} = 0$. Therefore, bound states can be found as energies $E$ and wavevectors $\Delta$ that satisfy the criteria defined in Eq.\,(\ref{eq:Constraints1}) and that ensure $T_{\alpha \alpha} = 0$, which can be found numerically.

\bibliography{supplement}